\documentclass[nofootinbib,preprint]{revtex4-1} 

\usepackage{amsmath}  
\usepackage{amsfonts} 
\usepackage{graphicx} 

\begin{document}



\title{The infalling photon, the infalling particle, and the observer at rest near the horizon of a black hole}

\author{Francisco D. Mazzitelli}
\affiliation{Centro At\'omico Bariloche and Instituto Balseiro, 
Comisi\'on Nacional de Energ\'\i a At\'omica, 8400 Bariloche, Argentina.}


\begin{abstract} 
When a massive test particle or a photon fall radially into a black hole, their energy, as measured by a static observer located very close to the horizon, diverges. In introductory courses on General Relativity,  this fact gives rise to questions about the reality of this divergence, and its eventual effect on the geometry of the black hole. We address these concerns and show that, eventually, it is the observer at rest who may induce corrections to the metric, unless its mass is crucially small when located near the horizon.  
\end{abstract}

\maketitle 

\section{Introduction}\label{sec:intro}

The Schwarzschild geometry is a static spherically symmetric solution of the vacuum Einstein's equations. In standard (or Schwarzschild) coordinates  it is 
given by \cite{booksGR}
\begin{equation}
ds^2=  - c^2(1-\frac{2MG}{c^2r})dt^2 +\frac {dr^2}{(1-\frac{2MG}{c^2r})}+ r^2(d\theta^2+\sin^2\theta d\phi^2)
\end{equation}
(from now on we will set the velocity of light $c=1$). Despite its simple appearance, there are no elementary derivations based on Newtonian gravity and special relativity \cite{imposs};  to derive it, one has to consider the full Einstein's  equations. The degree of difficulty in solving these equations depends crucially on the coordinates chosen \cite{stupid}.
 
As is evident the metric has a singularity at the Schwarzschild radius $R_S=2GM$, that defines the event horizon, but this turns out to be a problem associated to the coordinates.  Schwarzschild coordinates are valid 
and describe a spherically symmetric black hole as long as one considers $r>R_S$. The singularity of the coordinates at $r=R_S$ has some consequences, as a huge redshift of photons emitted by a source that is near the horizon and observed far from the black hole, a hint of the fact that the proper time for observers at infinity, $t$, is not an adequate coordinate to describe the geometry near the horizon. Traditional alternatives discussed in textbooks are Eddington-Finkelstein (EF) and Kruskal-Szekeres (KS) coordinates. Other interesting option 
is to use  Painlev\'e -Gullstrand coordinates \cite{regular}.

Let us consider an observer at rest (OAR)  
at a given radius $r=R$ (coordinates $\theta$ and $\phi$ are also fixed). Assume that a photon is emitted far from the horizon with a frequency $\omega_\infty$. The OAR will determine its frequency as $\omega^{(OAR)}_{IP}=\omega_\infty/\sqrt{1-R_S/R}$, and suffers a very large blueshift (the subindex IP stands for infalling photon). The frequency 
measured by the static observer diverges  as $R\to R_S$. In the same way,  the velocity of an infalling particle that is initially at rest far from the horizon, measured by the OAR,    
is given by $v_R=\sqrt{R_S/R}$ and tends to the speed of light as $R\to R_S$. Its energy diverges in this limit (these results are reviewed in the next section).

In introductory courses on General Relativity the students wonder whether this huge energy is real: Could this energy kill the rest observer? Can this energy be used? Why this energy does not modify Schwarzschild geometry? Or does it? The first two questions have immediate trivial answers: it is the energy {\it measured}  by a rest observer, and therefore it is real and can be used as any other type of energy by the OAR. The last two questions could be rephrased in terms of the eventual  backreaction of the test particle over the metric. As we will see, their answers pinpoints the advantages of using different sets of coordinates,  and shed light on some conceptual issues related to the measurements performed by different observers near the horizon. 

There is an analogous situation in flat spacetime that illustrates the potential problem: in a given coordinate system,  assume there is a test point-particle 
\cite{point}
of mass $m$ which is at rest. For sufficiently small mass, spacetime is almost flat. Outside the particle,  the metric 
is given by the Schwarzschild metric with mass $m$. Consider now  a second particle with the same mass but moving with velocity $v$
with respect to the first one. Its energy-momentum tensor will be given by 
\begin{equation}
T^{\mu\nu}(t,\vec x)=\frac{m }{u^0}\delta^{(3)}(\vec x-\vec x_P(t))u^\mu u^\nu=\frac{m}{\sqrt{1-v^2}}\delta^{(3)}(\vec x-\vec x_P(t))\frac{dx^\mu}{dt}\frac{dx^\nu}{dt}\, ,
\end{equation}
where $\vec x_P(t)$ denotes the trajectory of the moving particle and $u^\mu=dx^\mu/d\tau$  the components of its four-velocity
(the explicit  form of the energy-momentum tensor for a point-particle can be derived from its action, see Eqs.(20)-(22) below). 
Note that, for any fixed value of $m$, no matter how small, the components of the energy-momentum tensor become bigger and bigger as $v\to 1$, and the ultrarelativistic particle could modify the spacetime metric (i.e. it could produce a strong gravitational field as measured by an observer 
comoving with the first particle).  For instance, in the limit $m\to 0$ and $v\to 1$, with   $mv/\sqrt{1-v^2}\to p$,  the spacetime is described  by a shockwave with nonvanishing curvature on the plane that moves with the particle and is  perpendicular to its velocity (this is the well-known Aichelburg-Sexl geometry \cite{AS}). In this system of coordinates, the curvature produced by the particle at rest is of order $m\, G$,
while the one produced by the moving particle is of order $p\,  G$, that could be much larger
(there are interesting geometrical effects produced by the shockwave \cite{DrayThooft}).
Will the infalling particle or the infalling photon produce a similar modifications to the geometry of the black hole?

The article is organized as follows. In Section II we review the  computation of  the frequency of the IP and of the energy of
the free falling particle as measured by the OAR. In Section III we consider the point of view of a free falling observer (FFO).
 We compute  the frequency of the IP in this reference frame.  We first consider the case
 in which the IP and the FFO meet at  $r>R_S$,  and  then generalize the results to arbitrary values of $r$.  In the latter case, the use of  
 coordinates that are regular at the horizon becomes crucial, and the calculation is done in EF coordinates.
 In Section IV we discuss the eventual backreaction on the geometry produced by either the FFO, the IP, or the OAR. To do this,
 we compute the components of the energy-momentum tensor for each of them,   near the horizon. Once more, the use of regular
 coordinates is compelling, and the calculation is performed in KS coordinates. We show that, in the limit $r\to R_S$,  the components of the energy-momentum tensor remain bounded  for both the FFO and the IP (the latter case is discussed in the  Appendix).
 On the contrary, for the OAR,  the energy-momentum 
 tensor is divergent in this limit.
 Section V contains the conclusions of our work: neither the FFO nor the IP  will  induce modifications to the black hole geometry, 
 as long as the rest mass of the FFO and the initial energy of the IP are much smaller than the black hole's mass. However, the OAR
 will produce a backreaction, unless its mass tends to zero as $R\to R_S$.

\section{The observer at rest's point of view}

As a warm up, we remind the reader the well known OAR's point of view,
discussed in many textbooks.   
Since the OAR has a fixed coordinate $r=R$, its four-velocity is given by
\begin{equation}
{\mathbf u_{OAR}}=(\frac{dt}{d\tau},0,0,0)=\frac{1}{\sqrt{1-\frac{R_S}{R}}}(1,0,0,0)\, .
\end{equation}
The energy of an object with four-momentum $\mathbf p$ as measured by the OAR is given by $E^{(OAR)}=-\mathbf p\cdot\mathbf u_{OAR}$. 

Let us first consider the IP. As it is a massless particle, its four-momentum 
$\mathbf p_{IP}$ satisfies $\mathbf p_{IP}\cdot \mathbf p_{IP}=0$. Moreover,
the metric is static, which implies that $g_{0\alpha}p_{IP}^\alpha=g_{00}p_{IP}^0$
is a constant, that we identify with (minus) the energy at infinity, that is, 
$\hbar\omega_\infty$. These two conditions imply 
\begin{equation}
\mathbf p_{IP}=\hbar \omega_\infty(\frac{1}{1-\frac{R_S}{R}}, -1,0,0)\, ,
\end{equation}
and therefore, when $r=R$:
\begin{equation}\label{blueshift}
\omega_{IP}^{(OAR)}=\frac{E_{IP}^{(OAR)}}{\hbar}= -\frac{1}{\hbar}\mathbf p_{IP}\cdot\mathbf u_{OAR} =\frac{ \omega_\infty}{\sqrt{1-\frac{R_S}{R}}}\, ,
\end{equation}
which is the standard result for the gravitational redshift (blueshift in this case). 

Let us now consider the infalling particle of rest mass $m$, that from now on will be named the FFO. If the FFO starts falling at rest from $r\to \infty$, its four-momentum reads
\begin{equation}\label{pffo}
\mathbf p_{FFO}=m \, \mathbf u_{FF0}=m\, (\frac{1}{1-\frac{R_S}{r}}, -\sqrt{\frac{R_S}{r}},0,0)\, .
\end{equation}
This can be derived from $\mathbf p_{FFO}\cdot\mathbf p_{FFO}=-m^2$ and the conservation of energy $g_{00}p_{FFO}^0=-m$. 
As a consequence, the energy of the FFO measured by the OAR is given by
\begin{equation}
E_{FFO}^{(OAR)}=-\mathbf p_{FFO}\cdot\mathbf u_{OAR}=\frac{m }{\sqrt{1-\frac{R_S}{R}}}\, ,
\end{equation}
that corresponds to a relativistic particle with velocity $v_R=\sqrt{R_S/R}$. The velocity of the FFO measured by the OAR tends to $1$,  and the energy diverges,  as $R\to R_S$.

\section{The free-falling observer's view}  

We now consider the point of view of the FFO. We will first compute the energy of the IP as measured by the FFO
assuming that they meet at $r>R_s$,  and that the FFO is initially at rest at infinity. Then, we will generalize the calculation 
for arbitrary values of $r$ and of the initial energy of the FFO.   

If the FFO and the IP meet at $r>R_S$, a simple way to compute the energy of the IP measured by the FFO
is to realize that the measurements of the FFO and of the OAR  at $r=R$ can be connected by a Lorentz transformation \cite{Taylor}, a boost with velocity $v_R$. Therefore,  the energy of the IP measured by the FFO, $\omega_{IP}^{(IO)}$, is related to the energy measured by the OAR by a Doppler shift factor $\sqrt{(1-v_R)/(1+v_R)}$
\begin{equation}\label{IP-FFO}
\omega_{IP}^{(FFO)}=\omega_{IP}^{(OAR)} \sqrt{\frac{1-v_R}{1+v_R}}= \frac{\omega_\infty}{1+\sqrt{\frac{R_S}{R}}}\, ,
\end{equation}
where we used Eq.\eqref{blueshift}  and $v_R=\sqrt{R_S/R}$. It is a simple and instructive exercise to obtain the same result by projecting the four-momentum of the IP on the four-velocity of the FFO, i.e. $\omega_{IP}^{(FFO)}=-\frac{1}{\hbar}\mathbf p_{IP}\cdot\mathbf u_{FFO}$.

Note that the energy of the photon is finite when measured by a FFO even if they meet at the horizon. This is a consequence of the cancellation between the gravitational blueshift and the Doppler redshift associated
with a relative velocity which is close to the speed of light \cite{Dopplernote}.
Using similar arguments, one can compute the energy and velocity of the OAR as measured by the FFO. Of course, there is no need for additional calculations,
because $v_R$ is the {\it relative} velocity between them, that is univocally
defined  by the scalar product $\mathbf u_{FFO}\cdot \mathbf u_{OAR}$ \cite{vrel}.  
The energy of the OAR, measured by the FFO, is indeed also infinite as $R\to R_S$.

We  will now generalize the above results for arbitrary valu es of the meeting point $r$. 
If the FFO and the IP meet beyond the horizon,  in order  to compute $\omega_{IP}^{(FFO)}$ it is necessary to use a different set of coordinates,
to avoid the singularity of the Schwarzschild coordinates at the horizon. We will use the EF coordinates, which trade
the coordinate $t$ by a new coordinate $v$ defined as \cite{HartleEF}
\begin{equation}
v=t + r + R_S \log\vert\frac{r}{R_S}-1\vert\, .
\end{equation}
In the new coordinates, the spacetime interval reads
\begin{equation}
ds^2=-(1-\frac{R_S}{r})dv^2+ 2 dv dr + r^2(d\theta^2+\sin^2\theta d\phi^2)\, .
\end{equation}
Note that the metric is regular at the horizon.

Light cones associated to radially infalling photons are defined by $dv=0$. Therefore $p_{IP}^v=0$ and the only non-vanishing contravariant component of
the four-momentum  is $p_{IP}^r$. As the metric does not depend on the coordinate $v$, $g_{v\alpha}p_{IP}^\alpha=p_{IP}^r$
turns out to be constant, i.e. (minus) the energy of the IP at infinity, $\hbar\omega_\infty$. 

The four-velocity of the FFO reads
\begin{equation}
\mathbf u_{FFO}=(\frac{dv}{d\tau},\frac{dr}{d\tau},0,0)\, ,
\end{equation}
 and the two non-vanishing components can be obtained from the conservation law
 \begin{equation}\label{cons}
 -e_\infty=g_{v\alpha}u_{FFO}^\alpha=-(1-\frac{R_S}{r})\frac{dv}{d\tau}+\frac{dr}{d\tau}
 \end{equation}
 and the normalization
 \begin{equation}\label{norm}
 \mathbf u_{FFO}\cdot  \mathbf u_{FFO}=-1=-(1-\frac{R_S}{r} )\left(\frac{dv}{d\tau}\right)^2+ 2 \frac{dv}{d\tau}\frac{dr}{d\tau}\, .
 \end{equation}
 The constant $e_\infty\geq1$ is the energy per unit mass of the FFO, and equals $1$ when it is  initially at rest.
 For the sake of completeness,  we will compute $\omega_{IP}^{(FFO)}$ for an arbitrary initial energy. 
 
 The frequency of the IP, measured by the FFO , when they meet at $r$, is given by
\begin{equation}
\omega_{IP} ^{(FFO)}=-\frac{1}{\hbar}\mathbf u_{FFO}\cdot\mathbf p_{IP}=\omega_\infty\frac{dv}{d\tau}\, .
\end{equation}
We can compute $dv/d\tau$ by combining Eqs.\eqref{cons} and \eqref{norm}.  The combination gives the following quadratic equation for $dv/d\tau$
\begin{equation}
\left(1-\frac{R_S}{r}\right)\left(\frac{dv}{d\tau}\right)^2-2e_\infty\frac{dv}{d\tau}+1=0\, ,
\end{equation}
that
must be solved with the constraint $dr/d\tau<0$. The result is 
\begin{equation}
\frac{dv}{d\tau}=\frac{\omega_{IP}^{(FFO)}}{\omega_\infty}=\frac{1}{1-\frac{R_S}{r}}\left[e_\infty-\sqrt{e_\infty^2-1+\frac{R_S}{r} }\,\right ]\, ,
\end{equation} 
where $r$ is the radial coordinate at the meeting point. This expression is valid for all positive values of $r$ and tends to $1/(2e_\infty)$ as $r\to R_S$. When the FFO is initially at rest ($e_\infty=1$) we recover the previous result Eq.\eqref{IP-FFO} setting $r=R$. It is interesting to remark that $\omega_{IP}^{(FFO)}\to 0$ when the FFO and the IP meet near the black hole's singularity at $r\to 0$.

The fact that the FFO measures a finite value for the frequency of the IP at (and  beyond) the horizon, suggests that the IP should not introduce any backreaction
on the geometry. 

\section{The energy-momentum tensor in regular coordinates}

From  the FFO point of view, 
it is the OAR who has almost infinite energy. And viceversa. Then the question is: which of them, if anyone,   would produce a backreaction
on the geometry? 
Since we are interested in what happens near the horizon, it is again mandatory to use coordinates that are regular there. 

To evaluate the effect on the metric, one could solve Einstein's equations using the energy-momentum tensor with either the FFO, or the IP, or the OAR,  as a source, eventually
assuming that the metric describes a small perturbation of the Schwarzschild geometry in those coordinates. Instead of doing this, we will just evaluate the order of magnitude of the components of the energy-momentum tensor. We are admittedly using this very crude and simplistic approach,  estimating the backreaction through the magnitude of the source of Einstein's equations and neglecting radiation-reaction forces \cite{selforce}. This will be enough for  our purposes, which is to provide an elementary discussion in the context of introductory courses on General Relativity.

We will use KS coordinates $(T,X)$, that for $r>R_S$ are related with $(t,r)$ by \cite{HartleKS}
\begin{eqnarray}\label{KS}
T&=&\sqrt{\frac{r}{R_S}-1} \, e^{\frac{r}{2R_S}}\, \sinh[\frac{t}{2 R_S}]\nonumber\\
X&=&\sqrt{\frac{r}{R_S}-1} \, e^{\frac{r}{2R_S}}\, \cosh[\frac{t}{2 R_S}]\, .
\end{eqnarray}
The angular coordinates $(\theta,\phi)$ remain unchanged. In these coordinates, the spacetime interval reads
\begin{equation}
ds^2=\frac{4R_S^3}{r}\, e^{-\frac{r}{R_S}}(-dT^2+dX^2)+r^2(d\theta^2+\sin^2\theta d\phi^2)\, ,
\end{equation}
where $r$ should be considered as a function $r(T,X)$ defined implicitly by Eqs.\eqref{KS}. Indeed we have
\begin{equation}\label{implicit}
(\frac{r}{R_S}-1)\, e^{\frac{r}{R_S}}=X^2-T^2\, ,
\end{equation}
which shows that in KS coordinates the horizon is located at $X=T$ (this corresponds to $R\to R_S$ and $t\to\infty$
in Schwarzschild coordinates).  Note that,when $d\theta=d\phi =0$, the light cones are defined by $dT/dX=\pm 1$.

Assume that the mass $m$ of the FFO and of the OAR is small enough to be considered a test particle when
it is at rest far from the horizon. We will compute the components of the energy-momentum tensor associated to the mass,  using
 KS coordinates,  for both cases:  when the mass $m$ falls from very far (FFO) and when the mass is located at $R\simeq R_S$ (OAR). As already mentioned, these components should be used as a source in Einstein's equations to evaluate the backreaction of the particle on the metric.
If their values are always $O(m)$, we expect no appreciable backreaction on the metric. If some of the components blow up near the horizon, then the particle could modify the background geometry \cite{tech}.

The components of the energy-momentum tensor can be derived by taking the derivative with respect to the metric of the action for the point-particle:
\begin{equation}
S_P=-m\int d\tau\sqrt{-g_{\alpha\beta}u^\alpha u^\beta}
\end{equation}
and is given by \cite{Weinbergbook}
\begin{equation} \label{tmunuder}
T^{\mu\nu}(x)=\frac{2}{\sqrt g}\frac{\delta S_P}{\delta g_{\mu\nu}(x)}=m\int \frac{d\tau}{\sqrt {g(x)}}\delta^{(4)}(x-x_P(\tau)) u^\mu u^\nu, 
\end{equation}
where $u^\mu$ is the four-velocity of the particle and $x_P(\tau)$ its world line.  Performing the integral over proper time we obtain
\begin{equation}
T^{\mu\nu}(T, \vec x)=\frac{m}{\sqrt g}\delta^{(3)}(\vec x-\vec x_P(T))\frac{u^\mu u^\nu}{u^0}\, .
\end{equation}

As we already know the four-velocities of the observers in Schwarzschild coordinates, it is enough to consider the transformation rule for 
four-vectors
under a change of coordinates
\begin{equation}
u^{\alpha}=\frac{\partial x^\alpha}{\partial x^{'\beta}}u^{'\beta}\, .
\end{equation}
Here primed (non-primed)  coordinates  refer to Schwarzschild (KS) coordinates. For both the OAR and the FFO we will have
\begin{eqnarray}
u^T&=&\frac{\partial T}{\partial t}\frac{dt}{d\tau}+\frac{\partial T}{\partial r}\frac{dr }{d\tau}\nonumber\\
u^X&=&\frac{\partial X}{\partial t}\frac{dt}{d\tau}+\frac{\partial X}{\partial r}\frac{dr }{d\tau}\, .
\end{eqnarray}

For the OAR and by Eq.(3) we have $dr/d\tau=0$ and $dt/d\tau=1/\sqrt{1-R_S/R}$. Using the relation between KS and Schwarzschild coordinates, given in Eqs.\eqref{KS},  it is straightforward to obtain
\begin{eqnarray}\label{UOAR}
u_{OAR}^T&=&\frac{X}{2R_S}\frac{1}{\sqrt{1-\frac{R_S}{R}}}\nonumber\\
u_{OAR}^X&=&\frac{T}{2R_S}\frac{1}{\sqrt{1-\frac{R_S}{R}}}\, ,
\end{eqnarray}
where $R=R(X,T)$ is defined implicitly by Eq.\eqref{implicit}.

On the other hand, when the FFO meets the OAR at $r=R$, by Eq.(6) we have $dr/d\tau=-\sqrt{R_s/R}$ and $dt/d\tau=1/(1-R_S/R)$. Therefore,
\begin{eqnarray}\label{UFFO}
u_{FFO}^T&=&\frac{1}{2 R_S(1-\frac{R_S}{R})}\left(X-T\sqrt{\frac{R_S}{R}}\right)\nonumber\\
u_{FFO}^X&=&\frac{1}{2R_S(1-\frac{R_S}{R})}\left(T-X\sqrt{\frac{R_S}{R}}\right)\, .
\end{eqnarray}

Eqs.\eqref{UOAR} show that the components of the four-velocity
of the OAR
grow as the observer is located near the horizon ($X\simeq T, R\simeq R_S)$. Being proportional to $1/\sqrt{1-R_S/R}$, the components of the energy-momentum tensor grow accordingly.  To avoid this, the mass  of the OAR should be chosen smaller and smaller as $R\to R_S$. Note that in this case, the trajectory of the OAR is almost on the surface of the light cone:
\begin{equation}
\frac{dX}{dT}\big\vert_{OAR}=\frac{u_{OAR}^X}{u_{OAR}^T}=\frac{T}{X}\simeq 1\, .
\end{equation}

On the other hand, from  Eqs.\eqref{UFFO} one can show that the 
components of the four-velocity of the FFO, and therefore also those of its energy-momentum tensor,  remain finite as she/he crosses the horizon. To see this it is necessary to evaluate in detail the limit $R\to R_S$. Assume that $R/R_S=1+\delta$ with
$\delta\ll 1$. Taking into account that $X\to T$ as $\delta\to 0$, from Eq.\eqref{implicit} we have
\begin{equation}
X-T\simeq \frac{e\delta}{2 X}+ O(\delta^2)\, .
\end{equation}
Inserting this into Eqs.\eqref{UFFO} we obtain, to zeroth order in $\delta$, 
\begin{eqnarray}\label{UFFO2}
u_{FFO}^T&\simeq&\frac{X}{4 R_S}  \left(1+\frac{e}{X^2}\right) \nonumber\\
u_{FFO}^X&\simeq&\frac{X}{4R_S}   \left(1-\frac{e}{X^2}\right)  \, ,
\end{eqnarray}
and therefore
\begin{equation}\label{FFO3}
\frac{dX}{dT}\big\vert_{FFO}=\frac{u_{FFO}^X}{u_{FFO}^T}\simeq \frac{1-\frac{e}{X^2}}{1+\frac{e}{X^2}}\, ,
\end{equation}
whose absolute value is less than one for any finite value of the coordinate $X\simeq T$. 
These equations show that,  unlike the OAR, when  the FFO reaches the horizon the components of its four-velocity
remain bounded \cite{Augousti}.  Moreover,  its trajectory is inside the local light cone even when it is {\it on} the horizon (as it should).

It is  instructive   to check that the components of the energy-momentum tensor of the IP remain $O(\hbar \omega_\infty)$ as the photon crosses the horizon. The details are presented in the Appendix.

The results of this Section show that  neither the FFO nor the IP will induce modifications to the metric of the black hole. The mass of the OAR, instead, should  tend to zero as $R\to R_S$,   in order  to avoid the growth of the components of its energy-momentum tensor. 
 
\section{Discussion}

For a test particle, the energy-momentum tensor is proportional to its mass and,
as long as its components are $O(m)$ along the trajectory and $m$ is sufficiently small, it will not produce sizeable modifications to the metric.  Computing 
the energy-momentum tensor for the FFO in regular coordinates, we have seen that its components remain of order $m$ all along the trajectory. A similar argument applies for low-energy massless particles (see the Appendix below). Therefore, the infalling test particle  and the infalling low-energy photon will not produce corrections
to  the Schwarzschild geometry. On the other hand, for the OAR the components of the energy-momentum tensor diverge as $R\to R_S$, unless its mass is 
chosen to be smaller and smaller in this limit.  
In other words, if a small-mass spacecraft falls freely into a black hole, it will not modify significantly the metric. However, if at some point the engine is 
turned on and it remains at a fixed $R/R_S=1 + \delta$ with $\delta$ sufficiently small, it will. Even neglecting
the gravitational waves produced until  arriving at $r=R$.  An alternative observation that reinforces these results is the 
fact that  the proper acceleration of the OAR, given by  \cite{Doughty}
\begin{equation}
a_{OAR}=\sqrt {g_{\alpha\beta}\frac{du_{OAR}^\alpha}{d\tau}\frac{du_{OAR}^\beta}{d\tau}}=\frac{R_S}{R^2}\frac{1}{\sqrt{1-\frac{R_s}{R}}}\, ,
\end{equation}
also diverges as $\delta\to 0$.

The results we obtained may not be intuitive since, at first sight, one could assume that the FFO becomes
ultrarelativistic when approaching the horizon. 
Indeed, there seems to be  some confusion in the literature regarding the value of the velocity of the FFO when reaching the horizon.
While it is true that the OAR will determine that the velocity of the FFO is close to the velocity of light, this is not true
for all observers \cite{dispute},  as illustrated by Eq.\eqref{FFO3}. 
The strength of the gravitational field is also observer-dependent \cite{Mash}.

An analysis based on coordinates that are regular at the horizon 
shows that the ultrarelativistic  assumption for the FFO is wrong.  In fact,  in these coordinates
the path of the FFO lies well inside the light cone, while the path of the OAR is close to the light cone surface (the horizon!).
    
\section*{Acknowledgements} This research was supported by ANPCyT, CONICET, and UNCuyo. I would like to thank the International Centre for 
Theoretical Physics, Trieste, Italia, for hospitality during the initial stages of this work.

\section*{Appendix : The energy-momentum tensor  for the infalling photon}

In this Appendix we show that the components of the energy-momentum tensor for the IP remain finite  as the photon reaches the horizon of the black hole.  To begin with, we note that the energy momentum tensor for a massive particle in Eq.(22)  can be rewritten 
in terms of its four-momentum as
\begin{equation}
T^{\mu\nu}(T, \vec x)=\frac{1}{\sqrt g}\delta^{(3)}(\vec x-\vec x_P(T))\frac{p^\mu p^\nu}{p^0}\, .
\end{equation}
This expression is also  valid  for massless particles, as can be guessed from the fact that one can obtain the four-momentum
of a massless particle as the limit $m\to 0,  v \to 1$ of the four-momentum of a massive particle, assuming that
$m/\sqrt{1-v^2}$ remains finite in this limit,
\begin{eqnarray}
p_{m\neq 0}^0&=&\frac{m}{\sqrt{1-v^2}}\to p^0_{m=0}\nonumber\\
\vec p_{m\neq 0}&=&\frac{m\vec v}{\sqrt{1-v^2}}\to\vec  p_{m=0}\, .
\end{eqnarray}
 Therefore,  when the IP is located at $\vec x_{IP}(T)$ we have
 \begin{equation}
T^{\mu\nu}_{IP}(T, \vec x)=\frac{1}{\sqrt g}\delta^{(3)}(\vec x-\vec x_{IP}(T))\frac{p_{IP}^\mu p_{IP}^\nu}{p_{IP}^0}\, .
\end{equation}

We will compute
the components of the four-momentum of the IP near the horizon, using KS coordinates.  As for the OAR and the FFO,  we 
use the  transformation rule for four-vectors  under a coordinate transformation (see Eqs.(24)). In this case we have
\begin{eqnarray}\label{transp}
p_{IP}^T&=&\frac{\partial T}{\partial t}\, p_{IP}^t+\frac{\partial T}{\partial r}\, p_{IP}^r\nonumber\\
p_{IP}^X&=&\frac{\partial X}{\partial t}\, p_{IP}^t+\frac{\partial X}{\partial r}\, p_{IP}^r\,\,  .
\end{eqnarray}
The components of the four-momentum in Schwarzschild coordinates are given in Eq.(4).  Inserting this equation into
Eqs.\eqref{transp}, and using the relations between KS and Schwarzschild coordinates,  given in Eqs.(17), we obtain
\begin{equation}
p_{IP}^T=-p_{IP}^X=\frac{\hbar\omega_\infty}{2 R_S(1-\frac{R_S}{R})}(X-T)
\end{equation}
From  Eq.(28) we see that $X-T=O(1-R_S/R)$, and therefore the components of the four- momentum of the IP remain finite and  of  order $\hbar\omega_\infty$ as $R\to R_S$.
 The same happens with
the components of its energy-momentum tensor.


\begin{thebibliography}{99}
\bibitem{point}For simplicity we consider a ``point"-particle. By this we mean a particle whose size is much smaller than any other relevant scale in 
our discussion.  To be considered a test particle, its radius  ($R_P$) should be much larger than the Schwarzschild radius associated to the particle,
that is  $R_P\gg 2 m G$. Of course we will also assume $m\ll M$, the mass of the black hole.
\bibitem{Dopplernote} The factorization of the total frequency change into gravitational and kinematical (Doppler) terms has been discussed at length in the literature \cite{Doppler}.  Formulas that generalize Eq.\eqref{IP-FFO} for the case in which the two observers are in 
motion have also been derived \cite{ghosts},  and are relevant 
to analyze the communication between them in the vicinity of the horizon. Eq.\eqref{IP-FFO} will be enough for our discussion.
\bibitem{tech} A technical remark: although both KS  and EF coordinates are regular at the horizon,
for this discussion we prefer the use of KS coordinates, because all components of the metric are non-vanishing at the horizon. In this way,
it is possible to avoid concerns regarding which components  $T^{\mu\nu}, T^\mu_\nu$ or $T_{\mu\nu}$ should be computed.
\end{thebibliography}
\end{document}